\documentstyle[12pt,epsf]{article}
\begin{document}

\title
{Search for dibaryon production in $^4$Hep interactions at medium energies}
\author
{A. V. Blinov, M. V. Chadeyeva, and V. F. Turov\\ {\it Institute
for Theoretical and Experimental Physics (ITEP){\footnote{E-mail:
blinov@itep.ru, chadeyeva@itep.ru, turov@itep.ru}}},\\ {\it
Moscow, Russian Federation} }
\date{June 14, 2001}
\maketitle

\begin{abstract}
The  $^4$Hep interactions at 2.7~GeV/$c$ and 5~GeV/$c$  initial
momenta of $\alpha$-particles are investigated using the 2m liquid
hydrogen bubble chamber (the kinetic energies of the initial
protons in the nuclear rest frame were  $T_{\rm p} = 220$~MeV and
$620$~MeV). The effective mass spectrum of two nucleons from the
reactions  $^4$Hep$\rightarrow$dppn and $^4$Hep$\rightarrow$pppnn
was analysed. The narrow structure in the 2p mass spectrum from
the quasi-elastic reaction p$^4$He$\rightarrow$n$_{\rm F}$(pp)d at
$T_{\rm p} = 620$~MeV (where  n$_{\rm F}$ is the fast neutron in
the nuclear rest frame) was found which is the evidence of the
existence of dibaryon  with a mass $M_{2\rm{p}} = 2008 \pm 13$~MeV
and a width $\Gamma_{2\rm{p}} = 20 \pm 5$~MeV. The narrow
enhancements were also observed at the masses near the 2p
threshold $M_{2\rm{p}} = 1878 \div 1879$~MeV  in the reactions
p$^4$He$\rightarrow$n$_{\rm F}$(pp)d and
p$^4$He$\rightarrow$p$_{\rm F}$(pp)(nn).
\end{abstract}

1. INTRODUCTION

The question  whether the dibaryons exist or not is still the most intricate and
contradictory in the nuclear physics. Nowadays there are numerous theoretical as well
as experimental papers considering this problem (see, e.g. \cite{Tat99} where
the review of the  present-day experiments on the search of dibaryons  is
presented).  The overwhelming majority of the evidence for the narrow dibaryons
(with the width $\sim50$~MeV) was first obtained  in the experiments with the
use of bubble chambers. At the same time the results of some other experiments
performed in the same technique \cite{Kat84} as well as the electronic-based
ones (see \cite{Tat99} ) proved quite the opposite, i.e. they
showed  no evidence for such objects. Note most of the experiments for hunting the
dibaryons have been carried out in the course of  investigating the lepton and hadron
interactions with the lightest nuclei (d, $^3$He, $^4$He).
Earlier in the frameworks of the ITEP experiment for the study of nuclear
reactions in the few nucleon systems using  liquid-hydrogen bubble chambers
(LHBC) there was made a systematic search for the dibaryons in $^3$Hp- and
$^3$Hep- interactions at intermediate energies \cite{B85,A88-47,B88,A88-48}.
The mass spectra for NN and NN$\Delta$ systems
with the different isospin number projections were investigated. In the pp, pn, nn,
pn$\pi^+({\rm n}\Delta^{++})$ and pp$\pi^+({\rm p}\Delta^{++})$ mass spectra for the mass
region $1.88 \div 2.5$~GeV there was no evidence for dibaryon production.
This paper presents the search for dibaryon production in $^4$Hep- interactions
using the ITEP 2m LHBC. The momentum of the $^4$He nuclei is $2.7$~GeV/$c$ (the
kinetic energy of  initial protons in the nuclear rest frame is $T_{\rm p} =
220$~MeV) and $5$~GeV/$c$ ($T_{\rm p} = 620$~MeV). Note the evidence for
production of the narrow dibaryons in $^4$Hep- interactions became apparent for
the first time in the experiment with the JINR 1m LHBC exposed to the
$\alpha$-particles beam at 8.6~GeV/$c$ initial momentum \cite{Dubna83}.
Evidently the search for dibaryon in $^4$Hep- interactions seems to be more
attractive than for proton interactions with other lightest nuclei (d, $^3$H,
$^3$He) due to relatively smaller inter-nucleon ranges and respectively greater
probabilities for possible multi-quark bags with hidden color for $^4$He.

Earlier during the investigations of the effective mass spectra of two protons
in hadron-nucleus and NN interactions the narrow enhancements near the 2$m_{\rm
p}$ threshold were found \cite{Siem67,Azim74,Abr94}. They were treated in
\cite{Azim74,Abr94} as the result of the two-
proton final state interaction (the so-called Migdal-Watson effect). The 2p mass
spectrum near the threshold (similar in this region to the relative momentum
distributions) may be useful to obtain the 2p correlation functions in order to measure
the space-time range of the interaction region for particles emission in the nuclear
reactions \cite{Koo77,Bay81,Bud90}. In this paper we analyze in detail the
effective mass spectra of two protons  near the threshold (for $^4$Hep-
interactions).

\bigskip
 2. EXPERIMENTAL TECHNIQUE

The ITEP 2m LHBC  was exposed to separated beam of $\alpha$-particles at 2.7~
GeV/$c$ and 5~GeV/$c$. The chamber was situated in a magnetic field of 0.92~T.
The primary-beam background particles (mainly deuterons) were easily separated
from $^4$He nuclei by visual estimation of the track ionisation. About 60,000
pictures on the 2.7~GeV/$c$ beam and about 120,000 pictures on 5~GeV/$c$ beam
were obtained with an average of about 5-8 initial particles for the chamber
extension. About 18,000-19,000 events were measured at each initial momentum. A
more detail description of the experimental and data processing procedure used
in the present experiment can be found in \cite{A93}. Note the experimental
technique applied here permits to analyse the data in $4\pi$-geometry.

The total $^4$Hep- cross section has the standard form \cite{A93}
and is determined by the account of the number of interactions in
the fiducial volume. The total cross section is equal to $109.4
\pm 1.8$~mb and $121.5 \pm 2.9$~mb at 2.7 and 5~GeV/$c$
respectively (the errors are statistical only). The systematic
error in the absolute normalisation of the cross section is $\sim
3\%$. The results presented here are based on the data of the
reactions

\begin{equation}
  ^4{\rm Hep} \rightarrow {\rm dppn}  \label{eq1}
\end{equation}
\begin{equation}
  ^4{\rm Hep} \rightarrow {\rm pppnn}  \label{eq2}
\end{equation}

For particle identification in the case of three-prong events we used the
selection procedure standard for bubble chamber experiments taking into account the
secondary track ionization measurements. The events of the reaction
(\ref{eq1}) with only one neutral particle in the final state  make the
kinematics
balance possible. The events of the reaction (\ref{eq2}) with two neutral
particles are unbalanced. Note that at 2.7~GeV/$c$ ($T_{\rm p} =
220$~MeV) below the pion production threshold in the elementary NN process the
pion production in $^4$Hep- interaction is negligible.

\begin{figure}[p]
\vskip -2cm \epsfysize=20cm \centerline{\epsffile{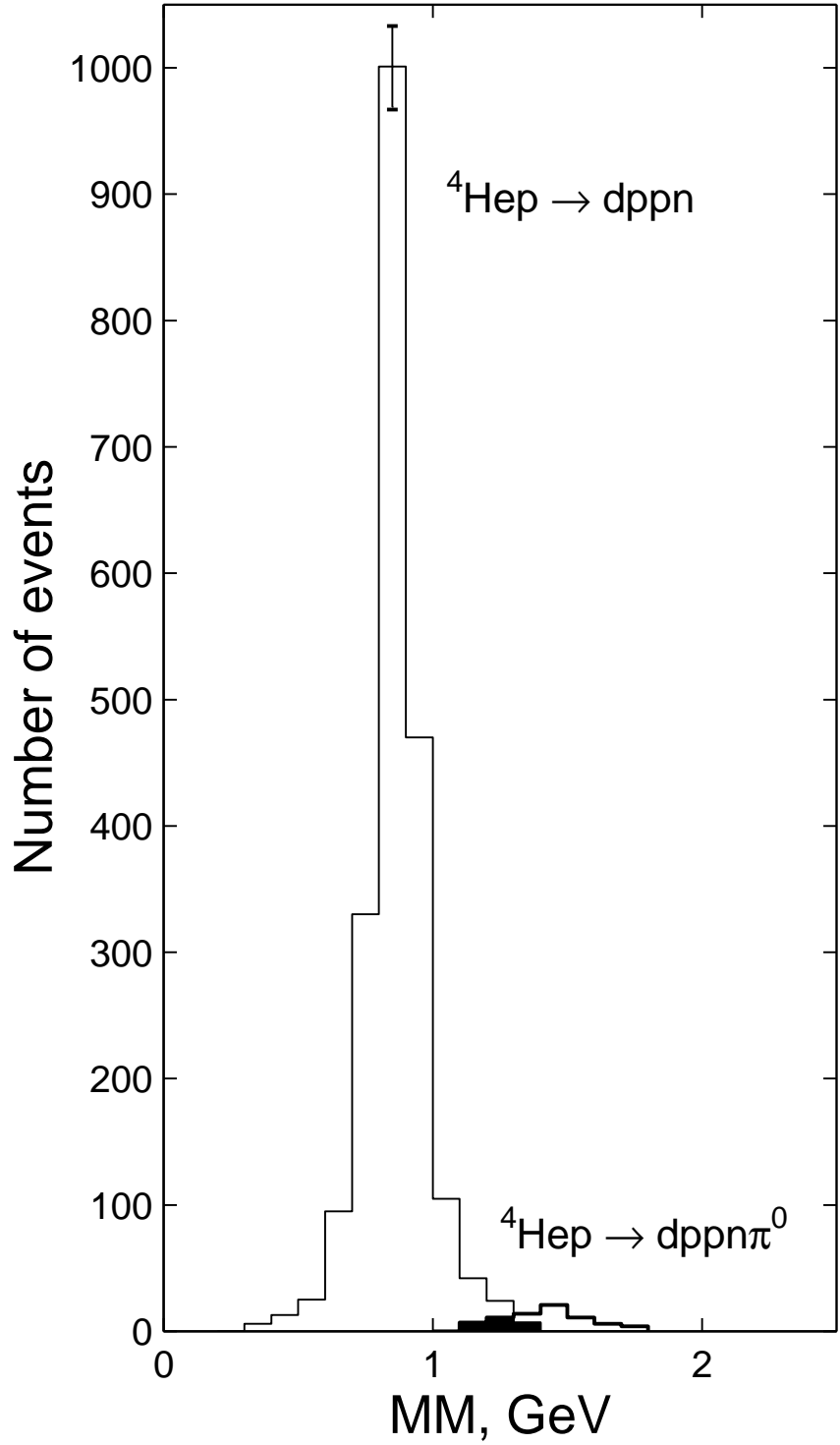}}
\caption{\small The missing mass squared $MM^2$ distributions for
unbalanced events of the reaction $^4$Hep$\rightarrow$dppn (thin
histogram) and for the ones of the reaction
$^4$Hep$\rightarrow$dppn$\pi^{0}$ (thick histogram) at $5$~GeV/$c$
initial momentum.} \label{fig1}
\end{figure}

The missing mass squared $MM^2$ distribution for unbalanced events
of the reaction (\ref{eq1}) as well as the one for the events of
the non-fitted reaction  $^4$Hep$\rightarrow$dppn$\pi ^0$ at
5~GeV/$c$ initial momentum is shown in Fig.\ref{fig1} (the
distributions presented here are based on the part ($\sim 80\%$)
of the total statistics). The $MM^2$ distribution for the reaction
\ref{eq1} has a gauss shape with the average close to the neutron
mass squared ${m_n}^2 = 0.88$~GeV$^2$. The overlap of the events
for the channels with and without neutral pion production (which
is marked by blacked domain in Fig.1) is $\sim 1\%$ of the total
number of events of the reaction (\ref{eq1}). The channel
(\ref{eq2}) at 5~GeV/$c$ evidently has some admixture of the
events of the reaction $^4$Hep$\rightarrow$pppnn $\pi ^0$ which is
$\sim 5\%$ to our estimate.

It is reasonable to sort out  the events of the reaction
(\ref{eq1}) into two classes with the secondary nucleons being the
fastest in the nuclear rest frame, i.e. the direct channel (where
protons are the fastest) and quasi-elastic charge exchange channel
(where neutrons are the fastest). The cross sections and the
number of events in each channel for two values of the  initial
momentum are presented in Table 1 (only statistical errors are
indicated). Note that in some events ($\sim 10\%$) deutrons are
the fastest in the nuclear rest frame. Such events with all the
three nucleon-spectators  are certainly  included in one or
another class in the analysis of effective mass distributions.

\begin{table}[tbh]
\caption
{\small Cross sections of the reactions $^4$Hep$\rightarrow$dppn and
$^4$Hep$\rightarrow$pppnn at $2.7$~GeV/$c$ ($T_{\rm p} = 220$~MeV) and
$5$~GeV/$c$ ($T_{\rm p} = 620$~MeV) initial momenta.}
\begin{center}
\begin{tabular}{|c|cc|p{2cm}p{2cm}|r@{$\pm$}l|}
\hline
 Momentum &  \multicolumn{2}{c|}{Channel} &
 \multicolumn{2}{c|}{Number of events} &  \multicolumn{2}{c|}{Cross section} \\
 (GeV/$c$) & \multicolumn{2}{|c|}{ } &  \multicolumn{2}{|c|}{ } &
\multicolumn{2}{c|}{(mb)} \\ \hline
 & & \multicolumn{1}{|c|}{Direct} & \multicolumn{1}{|c|}{2345} & &
\multicolumn{2}{|c|}{ } \\ [5pt] \cline{3-4}
\multicolumn{1}{|c|}{2.7} & $^4$Hep$\rightarrow$dppn
 & \multicolumn{1}{|c|}{Charge-} & \multicolumn{1}{|c|}{ } &
\multicolumn{1}{|c|}{3494} & 21.4 & 0.4 \\
 & & \multicolumn{1}{|c|}{exchange} &  \multicolumn{1}{|c|}{1149} & &
\multicolumn{2}{|c|}{ } \\ [5pt] \cline{2-7}
 & \multicolumn{2}{|c|}{$^4$Hep$\rightarrow$pppnn} &
\multicolumn{2}{|c|}{1620}  & 9.9 & 0.2 \\ [5 pt] \hline
 & & \multicolumn{1}{|c|}{Direct} & \multicolumn{1}{|c|}{1894} & &
\multicolumn{2}{|c|}{ } \\ [5pt] \cline{3-4}
\multicolumn{1}{|c|}{5} & $^4$Hep$\rightarrow$dppn
 & \multicolumn{1}{|c|}{Charge-} & \multicolumn{1}{|c|}{ } &
\multicolumn{1}{|c|}{2567} & 21.2 & 0.4 \\
 & & \multicolumn{1}{|c|}{exchange} &  \multicolumn{1}{|c|}{673} & &
\multicolumn{2}{|c|}{ } \\ [5pt] \cline{2-7}
 & \multicolumn{2}{|c|}{$^4$Hep$\rightarrow$pppnn} &
\multicolumn{2}{|c|}{1394}  & 11.5 & 0.3 \\ [5 pt] \hline
\end{tabular}
\end{center}
\end{table}

\bigskip
3.RESULTS AND CONCLUSIONS

\begin{figure}[p]
\vskip -2cm \epsfysize=18cm \centerline{\epsffile{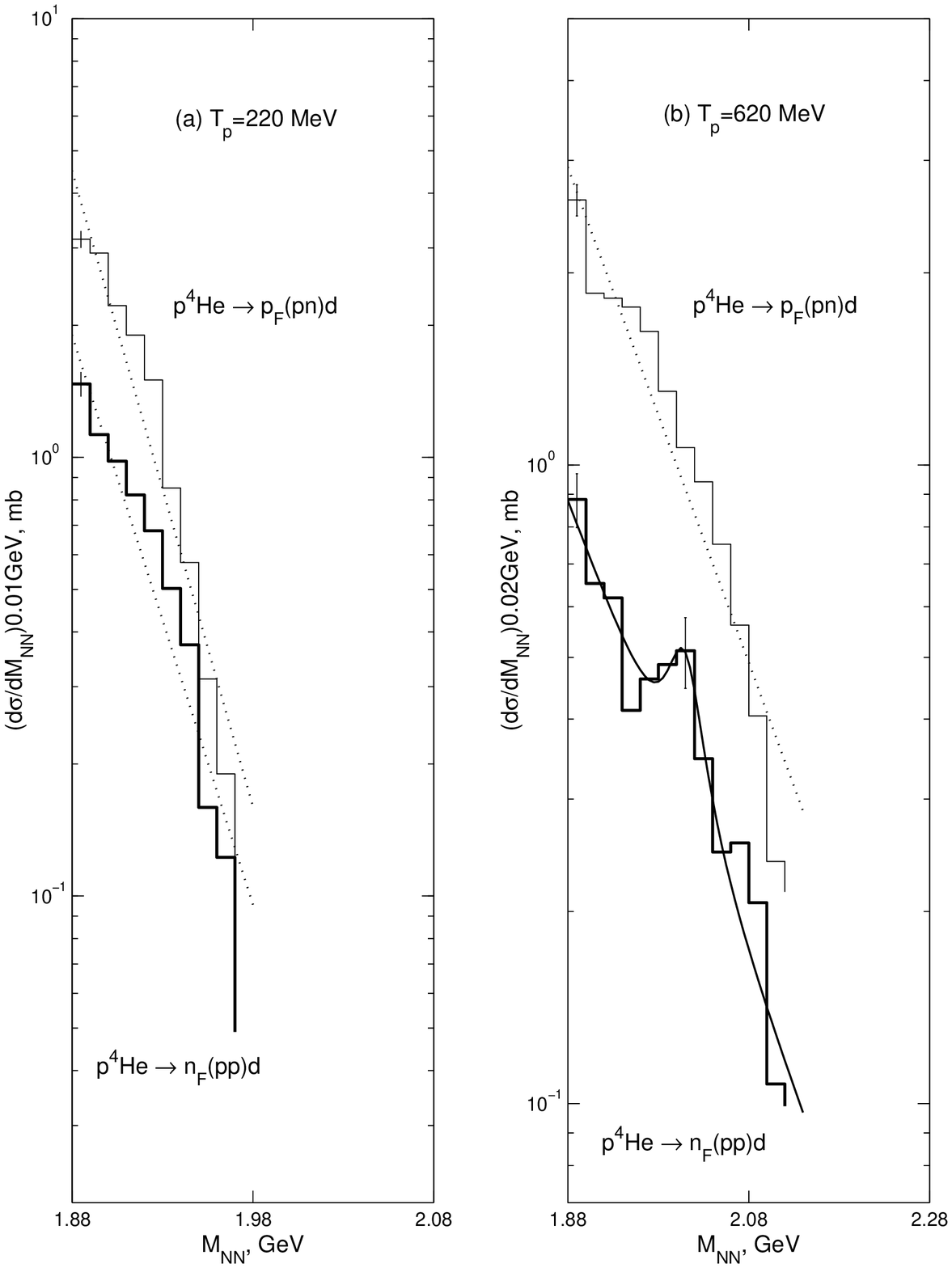}}
\caption{{\small The effective mass distribution of two
nucleon-spectators in the $^4$He rest frame for the direct channel
p$^4$He$\rightarrow$p$_{\rm F}$(pn)d (the upper histograms) and
quasi-elastic charge exchange channel p$^4$He$\rightarrow$n$_{\rm
F}$(pp)d (the lower histograms), where p$_{\rm F}$ (n$_{\rm F}$)
is the fast proton (neutron) in the $^4$He rest frame, at $T_{\rm
p} = 220$~MeV (a) and  $T_{\rm p} = 620$~MeV (b). The dotted lines
correspond to the exponential approximation of the data in the
intervals $1.88$~GeV $\le m_{\rm NN} \le 1.97$~GeV and $1.88$~GeV
$\le m_{\rm NN} \le 2.12$~GeV respectively. The solid line shows
the fit of the data in the interval $1.88$~GeV $\le m_{\rm NN} \le
2.12$~GeV by the sum of  Breit-Wigner function with the parameters
$M_{2\rm p} = 2008$~MeV and $\Gamma_{2\rm p} = 46$~MeV and the
exponential function for phase space (the dotted line in (b) in
this case corresponds to the exponential approximation of the data
out of the interval $1.96$~GeV $\le m_{\rm NN} \le 2.06$~GeV).}}
\label{fig2}
\end{figure}

Fig.\ref{fig2} shows the effective mass distribution of two
nucleon-spectators for the direct channel
p$^4$He$\rightarrow$p$_{\rm F}$(pn)d (the upper histograms) and
quasi-elastic charge exchange channel p$^4$He$\rightarrow$n$_{\rm
F}$(pp)d (the lower histograms), where p$_{\rm F}$ (n$_{\rm F}$)
is the fast proton (neutron) in the $^4$He rest frame, at $T_{\rm
p} = 220$~MeV (a) and $T_{\rm p} = 620$~MeV (b).

 In the two nucleon spectra for direct channel
as well as in the spectrum for charge exchange at $T_{\rm p} = 220$~MeV there are no
peculiarities. The dotted lines in Fig.\ref{fig2}a and Fig.\ref{fig2}b correspond to the
exponential approximation of the data in the intervals $1.88$~GeV $\le m_{\rm
NN} \le 1.97$~GeV and $1.88$~GeV $\le m_{\rm NN} \le 2.12$~GeV respectively.
In the two nucleon spectrum for charge exchange channel there is a narrow
structure (the enhancement is observed at the level of $3.1$ standard
deviations). The solid line in Fig.\ref{fig2} shows the fit of the data in the interval
$1.88$~GeV $\le m_{\rm NN} \le 2.12$~GeV by the sum of  Breit-Wigner function
with the parameters $M_{\rm 2p} = 2008 \pm 7$~MeV and $\Gamma_{\rm 2p} = 46 \pm
14$~MeV and the exponential function for phase space ($\chi ^2 / \rm{NDF} =
7.6/8$). The dotted line in Fig.\ref{fig2}b in this case corresponds to the
exponential approximation of the data out of the interval $1.96$~GeV $\le m_{\rm
NN} \le 2.06$~GeV. Note the fit of the data in the interval
$1.88$~GeV $\le m_{\rm NN} \le 2.12$~GeV by the exponential function only leads
to the value $\chi ^2 / \rm{NDF} =  21/10$ which does not permit to treat such
approximation as statistically significant.

The main results of the analysis of the experimental distributions in
Fig.\ref{fig2} are as follows.

1.The position and width of the maximum observed in the 2p mass spectrum
in the present experiment are near to the ones from \cite{R90} ($M_{\rm 2p} =
2007 \pm 15$~MeV, $\Gamma_{\rm 2p} = 39 \pm 17$~MeV) and from  other
experiments as well (see \cite{Tat99}).

2. Comparing the mass spectrum for charge-exchange channel at two energies
we may  conclude that the observed structure is apparently connected to the
manifestation of the non-nucleon degrees of freedom in the reaction considered (it is
particularly  indicated by  the nearness of the peak position to the summed masses of
the two nucleons and pion) and is not likely to be described on the basis of the
theoretical models considering the nucleon interaction mechanisms only (the multiple
scattering model, the pole model and so on).

3. If we assume that the observed structure is caused by the dibaryon
excitation with isospin I=1 it is no wonder that similar structures do not
appear in the direct channel as the background conditions for charge-exchange
reaction look more favorable. To determine the mass and width of the dibaryon
itself we use the modification of the usual Breit-Wigner function with the
account of the experimental resolution in the following form (see, i.e.
\cite{R90})

\begin{equation}
BW(m_{2\rm p}) = \frac{1}{\sqrt{2\pi}}\int \frac{BW(m)}{\sigma(m)}
\exp(-\frac{(m_{2\rm p}-m)^2}{2\sigma^2(m)}) {\rm d}m, \label{eq3}
\end{equation}

\noindent where $\sigma(m)$ is the experimental uncertainty  in
$m_{2\rm p}$. Thus we obtain the parameters of the assumed
dibaryon: the mass $M_{\rm 2p} = 2008 \pm 13$~MeV and the width
$\Gamma_{\rm 2p} = 20 \pm 5$~MeV ($\chi ^2 /{\rm NDF} = 7.9/8$) in
an excellent agreement with the results of \cite{R90} ($M_{\rm 2p}
= 2009 \pm 15$~MeV and $\Gamma_{\rm 2p} = 16 \pm 19$~MeV). The
dibaryon production cross section is $\approx 0.45 \pm 0.06$~mb.

\begin{figure}[p]
\vskip -2cm \epsfysize=20cm \centerline{\epsffile{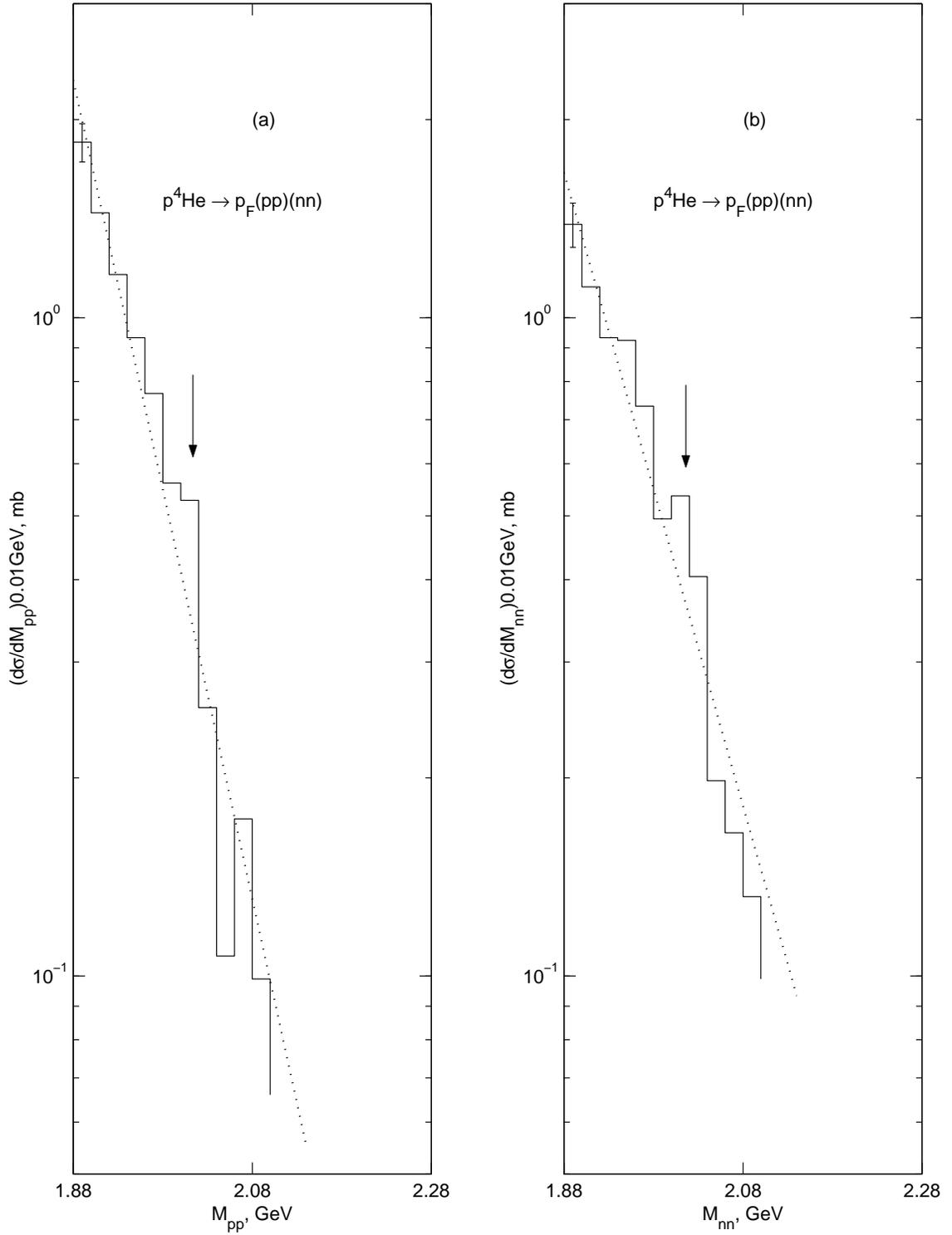}}
\caption{{\small The effective mass distribution of two
nucleon-spectators in the $^4$He rest frame for the
p$^4$He$\rightarrow$p$_{\rm F}$(pp)(nn) where p$_{\rm F}$ is the
fast proton in the $^4$He rest frame at $T_{\rm p} = 220$~MeV(a)
and $T_{\rm p} = 620$~MeV(b).The dotted lines correspond to the
exponential fit of the data in the intervals  $1.88$~GeV $\le
m_{\rm NN} \le 1.97$~GeV
 and  $1.88$~GeV $\le m_{\rm NN} \le 2.12$~GeV
 respectively.}} \label{fig3}
\end{figure}

In view to search similar structures in other $^4$Hep- interaction channels we
investigate the NN mass spectra in reaction (\ref{eq2}) where the summarised
statistics is sufficient for the analysis. Fig.\ref{fig3} shows the effective
mass distribution of two nucleon-spectators for the reaction
p$^4$He$\rightarrow$p$_{\rm F}$(pp)(nn) where p$_{\rm F}$ is the fast proton in
the $^4$He rest frame at $T_{\rm p} = 220$~MeV (a) and $T_{\rm p} = 620$~MeV
(b). In the mass spectra there are no peculiarities. A small enhancement of the
statistics over the exponential background (the dotted lines in Fig.\ref{fig3})
at $M_{\rm NN} \approx 2008$~MeV (marked by the arrow) is not
statistically significant.

\begin{figure}[p]
\vskip -2cm \epsfysize=12cm \centerline{\epsffile{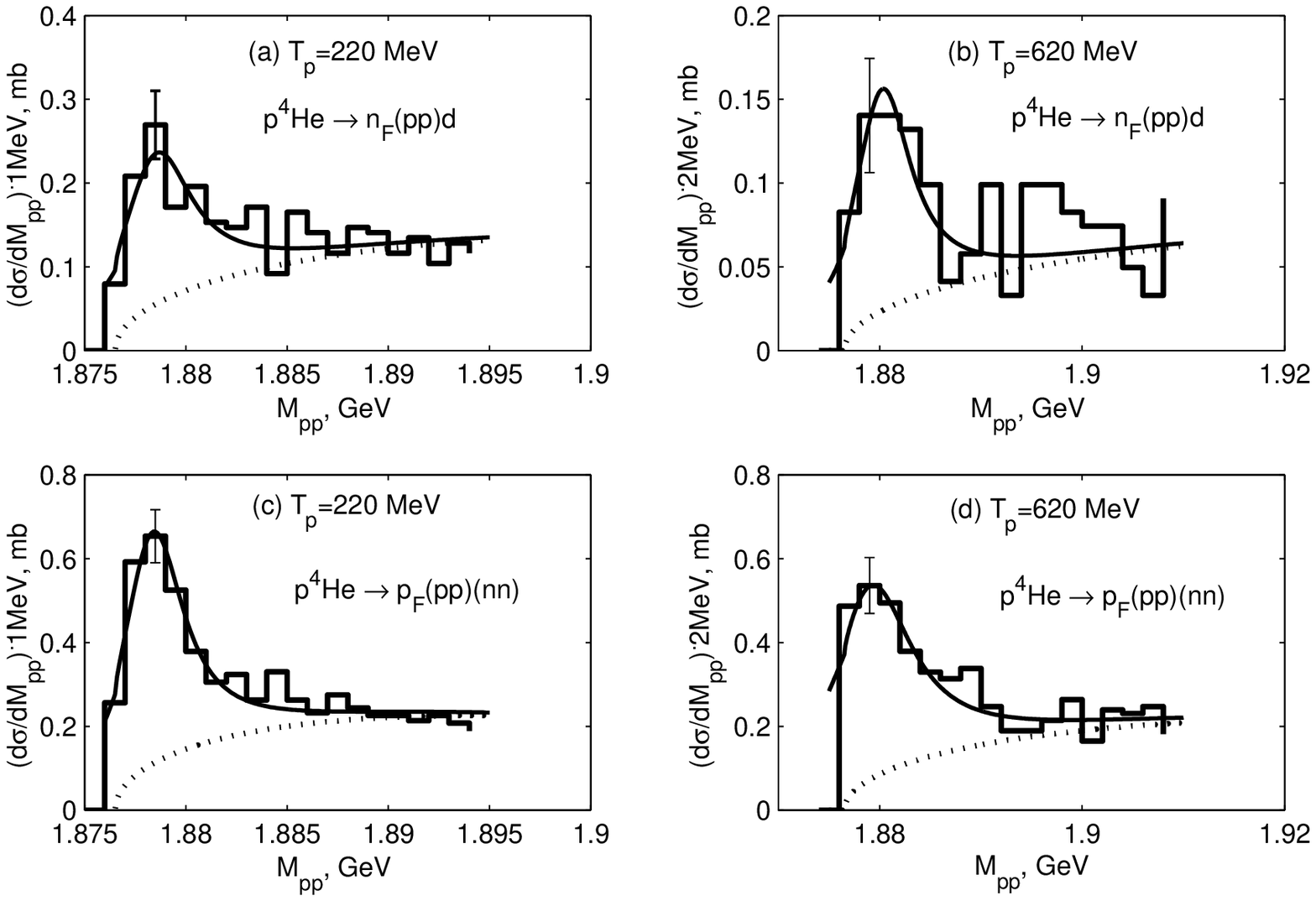}}
\caption{{\small The effective mass distribution of two
proton-spectators near $2m_{\rm p}$-threshold for the reaction
p$^4$He$\rightarrow$n$_{\rm F}$(pp)d at $T_{\rm p} = 220$~MeV (a),
$T_{\rm p} = 620$~MeV (b) and the reaction
p$^4$He$\rightarrow$p$_{\rm F}$(pp)(nn) at $T_{\rm p} = 220$~MeV
(c), $T_{\rm p} = 620$~MeV (d). The solid curves correspond to the
approximation of the data as the sum of the function which
describes the two-particle phase-space volume (see the text) and
Breit-Wigner function. The phase-space contribution is shown by
the dotted curves.}} \label{fig4}
 \end{figure}

    Earlier while studying the two-proton mass spectra in the reaction
p+n $\rightarrow$ p+p+$\pi^{-}$(backward) at $1.98$~GeV/$c$ momentum near
$2m_{\rm p}$-threshold there was observed a narrow enhancement at $1877.5 \pm
0.5$~MeV with the width  $2 \pm 0.5$~MeV \cite{Abr94}.
The effective mass distribution of two proton-spectators near
$2m_{\rm p}$-threshold for the reaction p$^4$He$\rightarrow$n$_{\rm F}$(pp)d at
$T_{\rm p} = 220$~MeV (a) , $T_{\rm p} = 620$~ MeV (b) and the reaction
p$^4$He$\rightarrow$p$_{\rm F}$(pp)(nn) at $T_{\rm p} = 220$~MeV (c) ,  $T_{\rm
p} = 620$~ MeV (d) is shown in Fig.\ref{fig4}. In the
mass region $1877 \div 1878$~MeV we may observe narrow (with the width $3 \div
5$~MeV) maxima. The solid curves in Fig.\ref{fig4} correspond to the
approximation of the data as the sum of the function $F_2$ and Breit-Wigner
function, where the function $F_2$ describes the two-particle phase space volume in
the following form (see. e.g. \cite{Book}):
\begin{equation}
F_2(m_{2{\rm p}}) = {\rm const} \cdot {(m_{2\rm p} - 2m_{\rm
p})}^{\frac{1}{2}} {( m_{max} - m_{2\rm p})}^{\alpha},  \label{eq4}
 \end{equation}
where for the reaction (\ref{eq1}) $\alpha = 2$,   $m_{max} = 2.0233$ and
$2.3304$
at $T_{\rm p} = 220$~MeV and $T_{\rm p} = 620$~MeV respectively,  for the
reaction (\ref{eq2}) $\alpha = 7/2$,  $m_{max} = 2.0211$ and $2.3282$ at
$T_{\rm p} = 220$~MeV and $T_{\rm p} = 620$~MeV respectively (only the
phase-space contribution (\ref{eq4}) is shown by the dotted curves). The
best fit of the data presented in Fig.\ref{fig4} in the form indicated above
gives the parameters of the observed peculiarity: $M_{2{\rm p}} = 1878.8 \pm
0.4$~MeV and $\Gamma_{2{\rm p}} = 4 \pm 1$~MeV ($\chi^2 / {\rm NDF} = 13.2/16$)
(see Fig.\ref{fig4}c).

As mentioned above the peculiarities observed in the pp mass spectra near
threshold may be of use to determine  the space-time range of the interaction region
for particles emission in the nuclear reactions. We hope to perform such analysis of
the data in the future.

    The main results of the paper are as follows.

(i) The narrow structure in the 2p-spectator mass spectrum from the quasi-
elastic reaction p$^4$He$\rightarrow$ n$_{\rm F}$(pp)d at $T_{\rm p} = 620$~MeV
(where n$_{\rm F}$ is the fast neutron in the nuclear rest frame) was seen which
is the evidence of the existence of dibaryon  with a mass $M_{2{\rm p}} = 2008
\pm 13$~MeV and a width $\Gamma_{2{\rm p}} = 20 \pm 5$~MeV. The enhancement is
observed at the level of $3.1$ standard deviations. The position and width of
the maximum observed in the present experiment are near to the ones from ref.
\cite{R90} and from other experiments as well (see \cite{Tat99}).

(ii) In the 2p mass spectra for the  reactions
p$^4$He$\rightarrow$n$_{\rm F}$(pp)d and p$^4$He$\rightarrow$ p$_{\rm
F}$(pp)(nn) we saw the narrow enhancements  at the masses
near the 2p threshold $M_{2{\rm p}} \approx 1878 \div 1879$~MeV which may be
useful to get information about the interaction region of the particles emission
in the $^4$Hep-interaction.

The authors are indebted to V.V.Kulikov, G.A.Leksin, V.V.Smolyankin and
A.V.Stavinskiy for useful critical comments and helpful suggestions.

\end{document}